\documentclass{sig-alternate-05-2015}

\usepackage{enumitem}
\usepackage{float}
\usepackage{enumerate}    
\usepackage{paralist} 
\usepackage{array}
\usepackage{multirow}
\newcolumntype{L}[1]{>{\raggedright\let\newline\\\arraybackslash\hspace{0pt}}m{#1}}
\newcolumntype{C}[1]{>{\centering\let\newline\\\arraybackslash\hspace{0pt}}m{#1}}
\newcolumntype{R}[1]{>{\raggedleft\let\newline\\\arraybackslash\hspace{0pt}}m{#1}}

\usepackage[table]{xcolor}
\usepackage{array,ragged2e}

\usepackage{flushend}

\begin{document}


\CopyrightYear{2018} 
\setcopyright{acmcopyright}
\conferenceinfo{CHASE'18,}{May 27 2018, Gothenburg, Sweden}





%

\title{Psychological Safety and Norm Clarity in Software Engineering Teams}

%
%
%
%
%

\numberofauthors{2} 
%
\author{
%
%
\alignauthor
Per Lenberg\\
       \affaddr{Div. of Software Engineering}\\
       \affaddr{Chalmers and University of Gothenburg}\\
       \email{perle@chalmers.se}
\alignauthor 
Robert Feldt\\
        \affaddr{Div. of Software Engineering}\\
        \affaddr{Chalmers and University of Gothenburg}\\
        \email{robert.feldt@chalmers.se}
}

\maketitle
\begin{abstract}
In the software engineering industry today, companies primarily conduct their work in teams. To increase organizational productivity, it is thus crucial to know the factors that affect team effectiveness. Two team-related concepts that have gained prominence lately are psychological safety and team norms. Still, few studies exist that explore these in a software engineering context.

Therefore, with the aim of extending the knowledge of these concepts, we examined if psychological safety and team norm clarity associate positively with software developers' self-assessed team performance and job satisfaction, two important elements of effectiveness.

We collected industry survey data from practitioners (N = 217) in 38 development teams working for five different organizations. The result of multiple linear regression analyses indicates that both psychological safety and team norm clarity predict team members' self-assessed performance and job satisfaction. The findings also suggest that clarity of norms is a stronger (30\% and 71\% stronger, respectively) predictor than psychological safety.

This research highlights the need to examine, in more detail, the relationship between social norms and software development. The findings of this study could serve as an empirical baseline for such, future work.
\end{abstract}


\printccsdesc



\section{Introduction}
The software industry has undergone a profound transformation in the last two decades, since the introduction of agile methods. These methods have shifted the focus from the individual developer and have instead highlighted teams and teamwork. Today, companies in the software industry mainly organize their work in teams. The team has replaced the individual as the critical organizational entity.

To increase productivity in software engineering organizations, it is, therefore, crucial to understand what factors most affect team and team members performance. Software engineering studies have, traditionally, focused on technological and process-oriented solutions to improve development~\cite{lenberg2014towards,lenberg2015behavioral}. However, in the last decades, researchers and practitioners alike have increasingly started to recognize that to improve team performance; one also needs to take human factors into account (e.g. ~\cite{beecham2008motivation,viana2012influence,Kosti2014EQ}). Software engineering researchers have, therefore, turned to the social sciences where the dynamics of groups have been studied for a long time. The insights gained lend themselves for use within the software engineering context and can be used to leverage performance in software teams. However, there is a need to study and add empirical evidence specifically for the software development context.

A psychological concept that has gained prominence among software practitioners the past years is psychological safety \cite{frazier2017psychological}. The concept is, commonly, defined as a shared belief among team members that the team is safe for interpersonal risk taking~\cite{edmondson1999psychological}. Psychological safety's widespread attention could, at least to some degree, be attributed to the description of Google's internal research results in high-profile publications and press articles~\cite{Mendoza2015how, duhigg2016google, delizonna2017HBR} in 2016.

According to these reports, Google conducted over 200 employee interviews and tested more than 250 concepts on well over 180 teams. These efforts resulted in a list of five main concepts that organizations need to consider when creating effective agile teams, i.e. psychological safety, dependability, structure and clarity, the meaning of work, and the impact of work. Of the five, psychological safety was identified as the far most important concept that underpinned the other four.

We recognize that the effects of psychological safety on team performance has a solid empirical foundation and that it repeatedly has been verified by researchers in several studies in the Social Sciences~\cite{newman2017psychological}. Despite the large body of support for the important role of psychological safety, only a few empirical studies exist that have analyzed its effect on software developers. Two exceptions are Faraj and Yan~\cite{faraj2009boundary} study from 2009 where they showed that boundary work is linked positively to psychological safety and Safdar et al.~\cite{safdar2017can} work from 2017 that explored how psychological safety affect software engineers selection of knowledge sources.

Another concept, related to psychological safety, that the social sciences repeatedly have found to strongly affect teams but that largely has been ignored by software engineering researchers, is social norms~\cite{lenberg2015behavioral, festinger1950social,levine1990progress, flynn2003s, salas2005there}. Norms are, commonly, defined as public or latent principles shared among the team members and that regulate and govern behavior~\cite{feldman1984development}. Researchers have demonstrated that \emph{what} norms teams adopt will affect the team performance, but the importance of making the norms known and understood by the team members have also been emphasized~\cite{wageman2005team}. Without such clarity of norms, team members do not realize what behaviors are accepted nor what responses to expect from others.

Only a few studies have explored norms in a software engineering context~\cite{lenberg2015behavioral}. Results from a small study by Teh et al.~\cite{teh2012social} revealed that software teams performed better when norms emphasized creativity rather than agreeability. Also, in an exploratory study, Stray et al.~\cite{stray2016exploring} conclude that team norms have an essential role in enabling team performance and suggest that teams regularly should reflect on their existing norms.

In this present study, we aimed to extend the knowledge and understanding of the two concepts \emph{psychological safety} and \emph{team norm clarity} in the software engineering domain. Therefore, we examined if these concepts associate positively with software developers' self-assessed \emph{team performance} and \emph{job satisfaction}, the latter being two widely known and significant factors in software organizations~\cite{dybaa2008empirical, dingsoyr2012team, francca2014motivated}.

We also sought to explore if raising \emph{team norm clarity} would strengthen the effect of \emph{psychological safety} on \emph{team performance} or \emph{job satisfaction}, i.e. if \emph{team norm clarity} moderated these relationships.

To meet these aims, we chose a quantitative approach. We collected industry data using questionnaires from five medium to large sized Swedish software engineering organizations (N = 217). All companies organized their work in teams and used an agile development approach. We analyzed the collected data using multiple linear regression analyses.

In the following section (\ref{sec_background}) we provide additional information regarding the concepts of team norms and psychological safety. Then, we detail the method used in this study (\ref{sec_method}) and, finally, the results are presented (\ref{sec_result}), discussed (\ref{sec_discussion}), and the paper concluded (\ref{sec_conclusion}).




\section{Background}
\label{sec_background}
In this section, we provide an overview of the concepts of team norms and psychological safety. 

\subsection{Team norms}
At work, we expect co-workers to behave in specific ways in particular circumstances. Each social situation entails its own set of expectations about the proper way to act. We, most of the time, conform to these guidelines that are provided by our organizational context and environment~\cite{sep-social-norms}. 

Furthermore, all group, over time, develop a set of rules, generally referred to as group norms, about how group members will interact with each other and perform their work. These group norms, defined as behavior patterns that are relatively stable and expected by group members~\cite{bettenhausen1991five}, are powerful and efficient mechanisms that regulate members' behavior~\cite{feldman1984development}.

Researchers commonly distinguish between descriptive or injunctive group norms~\cite{cialdini1991focus}. Descriptive norms are formed by observing what other group members do. They relate to how group members typically act, feel, and think in a given situation. Descriptive norms thus refer to the perception of \textbf{what is}. As an example, if you repeatedly notice that many team members arrive late to the daily stand-up, that would be a descriptive norm.

Injunctive norms, in turn, refer to the perception of \textbf{what ought to be}, i.e. what is approved or disapproved by other group members. For example, if you think that other team members would consider committing incomplete code as something that is wrong, that would be an injunctive norm. Injunctive norms develop through normative influence, or when group members conform to received social approval~\cite{deutsch1955study}. They influence by inducing rewards and sanctions for correct and incorrect behavior~\cite{ehrhart2004organizational}.

It has repeatedly been recognized that norms affect group performance~\cite{festinger1950social,levine1990progress, flynn2003s, salas2005there} and that the content of norms determines whether or not the efforts of group members will be directed at increasing the group's performance~\cite{ellemers2005identity}. Still, researchers understand relatively little about how and why specific norms emerge~\cite{flynn2003s}. 

Research suggests that group typically form their norms early in the lifespan of the group~\cite{ng2005antecedents}. The process of developing norms may be influenced heavily by forces of which members are unaware and may conflict with core management values about appropriate and expected group behavior~\cite{hackman1980work}. However, groups clearly cannot enforce norms covering every conceivable situation; instead, they are formed with respect to behaviors that have some significance for the group. According to Feldman et al.~\cite{feldman1984development}, a group will primarily enforce norms that facilitate its very survival, increase the predictability of group members' behaviors, and help the group avoid embarrassing interpersonal problems.

Wageman et al.~\cite{wageman2005team} and Hackman~\cite{hackman1980work} argue that productive teams hold core norms that actively promote continuous adaptation to the current situation and proactive planning strategies. Explicit specification of such core norms reduces the amount of energy the team members put on discussing acceptable behavior and facilitates the development of task performance strategies that are appropriate to the team's task and situation~\cite{wageman2005team}.

Despite the importance of group norms and the increased interest in behavioral aspects of software development, few studies have explored norms in a software engineering context~\cite{lenberg2015behavioral}. There are, however, a few exceptions. Results from a small study by Teh et al.~\cite{teh2012social} revealed that teams performed better when norms emphasized creativity rather than agreeability. The authors suggest norm manipulation as a practical way to enhance team performance. Also, in an exploratory study, Stray et al.~\cite{stray2016exploring} conclude that team norms have an essential role in enabling team performance and suggest that teams regularly reflect on both their injunctive norms and descriptive norms. To the best of our knowledge, no studies exist that explored the link between team norm clarity, team performance, and job satisfaction.

\subsection{Psychological Safety}
In the software industry, the developed systems are becoming increasingly sophisticated. That has emphasized the need for greater collaboration through activities such as information sharing and exchange of ideas among among employees and teams~\cite{edmondson2014psychological, newman2017psychological}. Even if such activities are beneficial for the company, they may come at a cost for the individual software engineer~\cite{detert2007leadership}.

For example, there is a fair chance that an implementation of a new idea, suggested by an employee, might not produce the expected improvements. If such idea is viewed as a failure by the organization, it portrays its inventor in a negative light. This type of organizational environment has the potential long-term consequence of hindering organizational development and learning~\cite{van1998helping, detert2007leadership, newman2017psychological}. One way to overcome such threats to the individual and organizational development and learning, is to create a safe work environment where employees feel comfortable to propose ideas, seek and provide honest feedback, take risks, and experiment~\cite{edmondson1999psychological}.

The concept of psychological safety, which is considered a social norm, was first explored by Schein and Bennis~\cite{schein1965personal} in their pioneering work on organizational change. Twenty-five years later the concept gained reviewed focus through William Kahn's seminal study on the relationship between psychological safety and personal engagement at work~\cite{kahn1990psychological}. More recently, Edmondson~\cite{edmondson1999psychological} has, yet again, highlighted the concept and proposed that psychological safety should be treated as a group-level concept estimating team climate. She defines psychological safety as a shared belief among employees as to whether it is safe to engage in interpersonal risk-taking in the workplace. There exist several other different definitions of the concept; however, the majority of the resent studies uses Edmondson's interpretation~\cite{newman2017psychological}.

Since Edmondson's publication at the turn of the century, the number of studies on psychological safety has increased dramatically~\cite{frazier2017psychological}. According to Newman et al.~\cite{newman2017psychological} there were, by the end of 2015, over 80 studies related to the concept of which the vast majority was empirical. Also, the past years there have been four publications that have reviewed and summarized the findings~\cite{sanner2013psychological, edmondson2014psychological, newman2017psychological,frazier2017psychological}. Together they conclude that the similarities in essential finding between studies are reliable and that the most prominent result is the strong relationship that psychological safety demonstrated with information sharing and learning behavior. Psychological safety is postulated as the critical foundation that enables behaviors essential to learning, whether the entity is an individual employee, a team, or an organization~\cite{frazier2017psychological}.

In more detail, research has, so far, identified several antecedents to psychological safety. There is growing support for a relationship between leadership behavior and psychological safety, where leadership related concepts such as support~\cite{may2004psychological} and trustworthiness~\cite{madjar2009trust} are central. Essential organizational concepts are employee perceptions of organizational support~\cite{carmeli2009relational}, access to mentoring~\cite{chen2014does}, and diversity practices~\cite{singh2013managing} foster psychological safety.


Moreover, studies have frequently shown that psychological safety affects various organizational outcomes. Numerous studies support its effect on the performance of individual employees and teams both directly~\cite{singh2013managing} and indirectly through facilitating learning behavior~\cite{sanner2013psychological}. In addition to performance, there is growing evidence of a link between employee perceptions of psychological safety and the organizational creativity and innovation~\cite{choo2004social, carmeli2010inclusive, kark2009alive, gu2013social}. At the individual level, several studies add support to a connection between psychological safety organizational commitment~\cite{chen2014does} and work engagement~\cite{may2004psychological}.

Despite the large body of support for the vital role of psychological safety, only a few empirical studies exist that have analyzed its effect on software development teams. None of these, however, explores the direct relationship between psychological safety and team performance.

In a study from 2009 that included 64 software development teams (N = 290), Faraj and Yan~\cite{faraj2009boundary} showed that boundary work, i.e. boundary spanning, boundary buffering, and boundary reinforcement, is positively linked to team performance and psychological safety, and that task uncertainty and resource scarcity moderated this effect.

Safdar et al.~\cite{safdar2017can} surveyed 1345 software engineering that provided substantial evidence that different levels of psychological safety affect how an individual software engineer will select one source of knowledge over others. Software engineers with high psychological safety levels were more inclined to consult fellow team member whereas individuals with low levels were more likely to choose external sources.

Finally, in a theoretical paper from 2017, Diegmann and Rosenkranz~\cite{diegmann2017team} propose a model and research design to investigate the effects of team diversity, psychological safety, and social agile practices on team resilience and team performance in agile software development.

\section{Method}
\label{sec_method}
As stated in the introduction, we aimed to examine if \emph{psychological safety} and \emph{team norm clarity} associate positively with \emph{team performance} and \emph{job satisfaction}. We also sought to explore if raising \emph{team norm clarity} would strengthen the effect of \emph{psychological safety} on \emph{team performance} or \emph{job satisfaction}.

To meet these aims, we collected industry data from five organizations using questionnaires. The data were processed using multiple linear regression analysis. In the following sections, we describe the sample, questionnaire, and the analysis.

\subsection{Sample}
We collected the data for this study from 38 software engineering teams working for five different organizations located in Sweden, see Table~\ref{table:companies}. Three of these were product development companies (1-3), while two were consultant companies (4,5). For commercial/NDA reasons, we cannot provide detailed descriptions of the companies.

All the participating companies stated that they use an agile development approach. Based on our knowledge of the organizations, which we gained through conversations with managers and a few development teams, we deem that the companies all were on level two or three on the Agile Adoption Framework developed by Sidky~\cite{sidky2007disciplined}. However, we did not measure this in detail.

\begin{table}[ht]
\footnotesize
\setlength\extrarowheight{2pt}
\centering
\begin{tabular}{| C{.09\textwidth} | C{.09\textwidth} | C{.09\textwidth}  | C{.09\textwidth} |} \hline
\rowcolor{black!15} Comp. Id & Comp. size & Resp. & Teams \\ \hline
1 & Large   & 114 (102)             & 20 (18) \\ \hline     
2 & Large   & 4                     & 1 \\ \hline           
3 & Medium  & 39                    & 7 \\ \hline           
4 & Large   & 6                     & 1 \\ \hline           
5 & Medium  & 54 (12)               & 9 (2) \\ \hline       
  &         & \textbf{217 (163)}    & \textbf{38 (29)} \\ \hline
\end{tabular}
\caption{Company size relates to the number of employees, where Large = more than 10 000 employees and Medium = more than 500 employees. Two departments in company 1 and 5 did not fill out the items related to \emph{job satisfaction}. The number for respondents for this concept is stated in parenthesis.}
\label{table:companies}
\end{table}

\subsection{Questionnaire}
\begin{figure}[htbp]
\centering
\includegraphics[width=0.48\textwidth]{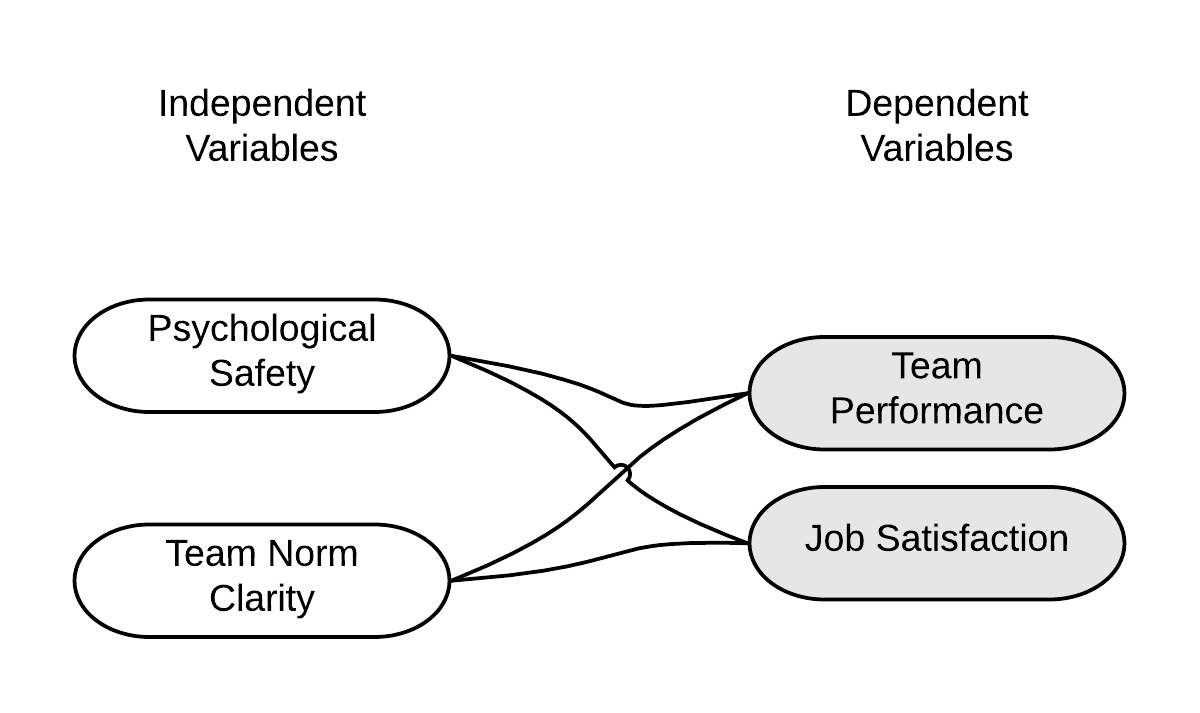}
\caption{Model used to examine if psychological safety and team norm clarity can predict team performance and job satisfaction.}
\label{fig_model}
\end{figure}

The questionnaire included twenty items related to the four variables in our simple and focused model presented in Figure~\ref{fig_model}. We aimed to utilize previously verified concepts rather than develop new scales and items. Therefore, \emph{job satisfaction} was measured using the four items defined by Thompson and Phua~\cite{thompson2012brief} and \emph{team performance} was estimated based on the items suggested by Cohen et al.~\cite{cohen1996predictive}.

To estimate \emph{psychological safety}, i.e. the first of two independent variables, we used the seven items defined by Edmondson~\cite{edmondson1999psychological}. It is, by far, the most commonly used measurement and it has been subjected to extensive validation tests, which have shown that the measure has strong content, criterion, and construct validity. The measurement has, also, been found to be reliable across diverse samples in various contextual settings~\cite{newman2017psychological}.

Finally, the second independent variable, i.e. \emph{team norm clarity}, was defined by Wageman, Hackman, and Lehman as a part of the team diagnostic survey~\cite{wageman2005team}. It included three items that collectively estimate if the team norms of conduct were clear and well accepted.

All items are presented in Table~\ref{table:items_model}. The answers to all of these items were measured using a five-point Likert scale with the following alternatives: 1 `strongly agree', 2 `agree', 3 `neither agree nor disagree', 4 `disagree', and 5 `strongly disagree'.

\begin{table}[ht]
\footnotesize
\setlength\extrarowheight{2pt}
\centering
\begin{tabular}{| C{.05\textwidth} | L{.36\textwidth} |} \hline
\rowcolor{black!15} \Centering Id. & \Centering Item  \\ \hline
PR1 & This team produce quality work. \\ \hline
PR2 & My team is productive. \\ \hline
PR3 & This team delivers according to schedule. \\ \hline
PR4 & This team delivers according to budget. \\ \hline
PR5 & My team communicates efficiently with others, e.g. product owner, line manager, and other teams. \\ \hline
PR6 & The team could become more efficient. (R) \\ \hline
JS1 & I find real enjoyment in my job. \\ \hline
JS2 & I like my job better than the average person. \\ \hline
JS3 & Most days I am enthusiastic about my job. \\ \hline
JS4 & I feel fairly well satisfied with my job. \\ \hline
NC1 & Standards for member behavior in this team are vague and unclear. (R)\\ \hline
NC2 & It is clear what is, and what is not, acceptable member behavior in this team.\\ \hline
NC3 & Members of this team agree about how members are expected to behave. \\ \hline
PS1 & If you make a mistake on this team, it is often held against you. (R) \\ \hline
PS2 & Members of this team are able to bring up problems and tough issues. \\ \hline
PS3 & People on this team sometimes reject others for being different. (R) \\ \hline
PS4 & It is safe to take a risk on this team. \\ \hline
PS5 & It is difficult to ask other members of this team for help. (R) \\ \hline
PS6 & No one on this team would deliberately act in a way that undermines my efforts. \\ \hline
PS7 & Working with members of this team, my unique skills and talents are valued and utilized. \\ \hline
\end{tabular}
\caption{The table shows the items used to compile the four variables; team performance (PR), job satisfaction (JS), team norm clarity (NC), and psychological safety (PS). If a question ends with (R), the response is to be inverted when creating the index variable.}
\label{table:items_model}
\end{table}

A confirmatory factor analysis (CFA) was used to test whether measures of the items were consistent with our understanding of that item's nature, and, as such, to test whether the data fit a hypothesized measurement model~\cite{fabrigar1999evaluating}. Due to limited space, we chose not to present the CFA table in this report. However, based on the CFA, one item from the psychological safety variable (item no. PS5 in Table~\ref{table:items_model}) and one item from the productivity variable (item no. PR6) were dropped and not included in the subsequent analysis.

Moreover, the Cronbach's alpha, a statistic calculated from the pairwise correlations between the items, are shown in table~\ref{table:Cronbach}. The estimates were somewhat lower than in previous research where they have been used, but still acceptable~\cite{tavakol2011making}.



\begin{table}[ht]
\footnotesize
\setlength\extrarowheight{2pt}
\centering
\begin{tabular}{| L{.20\textwidth} | C{.15\textwidth} |} \hline
\rowcolor{black!15} Concept & \Centering Cronbach Value \\ \hline
Psychological safety    &   .75 \\ \hline
Team norm clarity       &   .71 \\ \hline
Team performance        &   .71 \\ \hline
Job satisfaction        &   .86 \\ \hline
\end{tabular}
\caption{The Cronbach's alpha value for the included variables.}
\label{table:Cronbach}
\end{table}

\subsection{Data Collection Procedure}
We used two methods for distributing the survey. For company 1 and 5, we handed out the questionnaires ourselves on paper, earning us a response rate of approximately 80\%. However, for the other companies, i.e. 2, 3, and 4, we distributed the questionnaire electronically using Google Forms. That, unfortunately, resulted in a considerably lower response rate, roughly 30\%. The data collection was conducted during a five-months period, starting in November of 2016. 

Before filling out the survey, the participants were informed about the purpose of the research, that it was anonymous, that it was voluntary to participate and, finally, that the researcher would not share the raw or analyzed data with other researchers or with their respective management. We gave the participating teams the opportunity to get information about their team's results in relation to the overall mean of the other teams. Roughly half of them were interested in this feedback, which was provided by a researcher in May and June of 2017.

\subsection{Quantitative Data Analysis}
To test the proposed model, two multiple linear regression analyses were conducted with \emph{team performance} and \emph{job satisfaction} as dependent variables, while \emph{psychological safety} and \emph{team norm clarity} were independent variables. We followed the analysis procedures outlined by Meyers~\cite{meyers2006applied}.

In addition, using the method describe by Dawson~\cite{dawson2014moderation}, we tested if \emph{team norm clarity} moderated the relationship between \emph{psychological safety} and the two dependent variables \emph{team performance} and \emph{job satisfaction}.


Before conducting the analysis, we verified that the collected data actually could be analyzed using linear regression. A visual analysis of scatter plots for all variables indicated a linear relationship. Further, we checked the homoscedasticity and normality of residuals with the Q-Q-Plot. The plot indicated that in our multiple linear regression analysis there is no tendency in the error terms. Regarding autocorrelation, the Durbin-Watson values for the model were d = 1.88 (\emph{team performance}) and d = 2.19 (\emph{job satisfaction}). These are between the two critical values of 1.5 $<$ d $<$ 2.5 and, therefore, we can assume that there is no first order linear auto-correlation in our multiple linear regression data. Finally, the data were analyzed to determine the presence of multicollinearity. The variance inflation factors~\cite{o2007caution} were all well below three (VIF = 1.25 for \emph{team performance} and VIF = 1.27 for \emph{job satisfaction}), indicating only a small risk for multicollinearity.

The analysis was conducted using SPSS version 24.

\section{Result}
\label{sec_result}
We used two separate standard multiple linear regression analyses (N = 217 and N = 163) to examine if \emph{psychological safety} and \emph{team norm clarity} could predict \emph{team performance} and \emph{job satisfaction}. The correlations of the variables, shown in Table \ref{table:Correlation}, were moderate with a maximum value of .506 for \emph{job satisfaction} and \emph{team norm clarity}.

\begin{table}[htbp]
\footnotesize
\setlength\extrarowheight{2pt}
\centering
\begin{tabular}{| C{.12\textwidth} | C{.08\textwidth} | C{.08\textwidth} | C{.10\textwidth} |} \hline
\rowcolor{black!15} Variable    & Team norm clarity & Team performance    & Job satisfaction  \\ \hline
Psychological safety            &   .446**          & .389**                & .416**            \\ \hline
Team norm clarity               &   -               & .429**                & .506**            \\ \hline
Team performance              &   -               &   -                   & .386**           \\ \hline
\end{tabular}
\caption{Pearson r correlations of the variables. '**' indicates that the correlation is significant at .01 level.}
\label{table:Correlation}
\end{table}


\begin{table*}[t]
\footnotesize
\setlength\extrarowheight{2pt}
\centering
\begin{tabular}{| C{.2\textwidth} | C{.1\textwidth} | C{.1\textwidth} | C{.1\textwidth} | C{.1\textwidth} |C{.1\textwidth} |C{.1\textwidth} |} \hline
\rowcolor{black!15} Model   & B     &  SE-b & Beta  & Pearson r & sr\textsuperscript{2} & Sig\\ \hline
(Constant)                  & 1.37  & .137  &       &           &                       & .000 \\ \hline 
Psychological safety**      & .248  & .067  & .246  & .389      & .048                  & .000 \\ \hline 
Team norm clarity**         & .274  & .058  & .319  & .429      & .081                  & .000 \\ \hline
\end{tabular}
\caption{The raw and standardized regression coefficients of the predictors together with their correlations with \emph{team performance}, their squared semi-partial correlations (sr\textsuperscript{2}) and the significance level. The symbol '**' indicates that the variable was significant at .01 level.}
\label{table:Regression_teameffectivness}
\end{table*}

\begin{table*}[t]
\footnotesize
\setlength\extrarowheight{2pt}
\centering
\begin{tabular}{| C{.2\textwidth} | C{.1\textwidth} | C{.1\textwidth} | C{.1\textwidth} | C{.1\textwidth} |C{.1\textwidth} |C{.1\textwidth} |} \hline
\rowcolor{black!15} Model   & B     &  SE-b & Beta  & Pearson r & sr\textsuperscript{2} & Sig\\ \hline
(Constant)                  & .995  & .177  &       &           &                       & .000 \\ \hline 
Psychological safety**      & .276  & .088  & .233  & .416      & .043                  & .002 \\ \hline 
Team norm clarity**         & .391  & .073  & .399  & .506      & .125                  & .000 \\ \hline
\end{tabular}
\caption{The raw and standardized regression coefficients of the predictors together with their correlations with \emph{job satisfaction}, their squared semi-partial correlations (sr\textsuperscript{2}) and the significance level. The symbol '**' indicates that the variable was significant at .01 level.}
\label{table:Regression_jobsatisfaction}
\end{table*}

The analysis showed that the regression models for both \emph{team performance} and \emph{job satisfaction} were statistically significant (F(2, 214) = 32.47, p \textless .001; F(2, 160) = 43.1, p \textless .001). The former accounts for 23\% of the variance (R\textsuperscript{2} = .233, Adjusted R\textsuperscript{2} = .226), while the latter accounts for slightly more, 30\% (R\textsuperscript{2} = .299, Adjusted R\textsuperscript{2} = .290). We deem that the explained variance, especially for the \emph{job satisfaction} model, is rather high compared to other studies in social science considering that they include only two dependent variables. This adds support that our hypothesized models are relevant approximations that capture important factors. However, it is clear that future work should include more independent variables in more complex models to try to account for more of the variance.

\begin{figure}[htbp]
\centering
\includegraphics[width=0.48\textwidth]{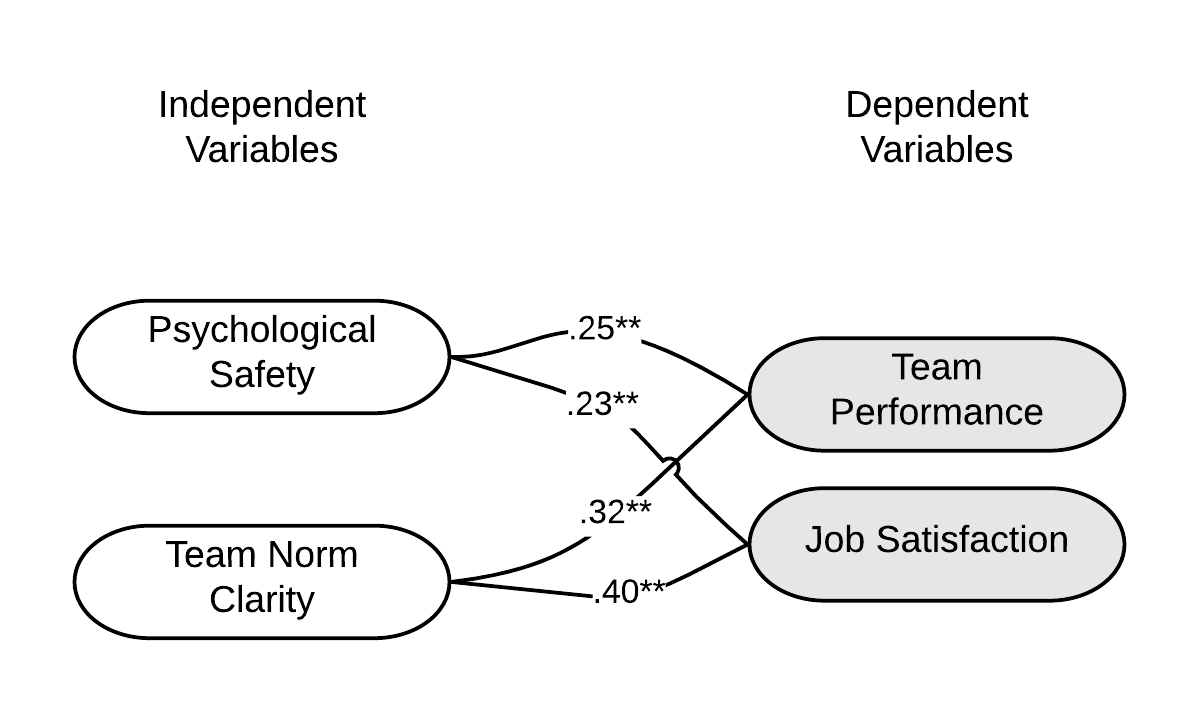}
\caption{Model used to predict \emph{team performance} and \emph{job satisfaction}. The connection line figures are the beta values from Table~\ref{table:Regression_teameffectivness} and \ref{table:Regression_jobsatisfaction}. '**' indicates that the value was statistically significant at .01. }
\label{fig:model_result}
\end{figure}

The results for the regression models are presented in Table~\ref{table:Regression_teameffectivness}, Table~\ref{table:Regression_jobsatisfaction}, and Figure~\ref{fig:model_result}. As can be seen, all variables have a significant result. However, with regard to the order of impact, \emph{team norm clarity} is a better predictor of team performance than \emph{psychological safety}, both in terms of contribution to the models (Beta) and also in terms of the unique variance it explains (indexed by the squared semi-partial correlation in column sr\textsuperscript{2}).

Based on this, \emph{team norm clarity} was examined as a moderator of the relation between \emph{psychological safety} and the two dependent variables \emph{team performance} and \emph{job satisfaction}. The interaction terms between \emph{team norm clarity} and \emph{psychological safety} did not explain a significant increase in variance for \emph{team performance} ($\Delta$R\textsuperscript{2} = .000, F(1, 218) = .005, p = .943) or \emph{job satisfaction} ($\Delta$R\textsuperscript{2} = .013, F(1, 164) = 2.92, p = .090). Thus, \emph{team norm clarity} was \textbf{not} a significant moderator of these relationships.

\section{Discussion}
\label{sec_discussion}
In this study, we examined if \emph{psychological safety} and \emph{team norm clarity} associate positively with \emph{team performance} and \emph{job satisfaction}. The result of two separate multiple linear regression analyses, where we used industrial data (N = 217 for \emph{team performance} and N = 163 for \emph{job satisfaction}), indicated that self-assessed \emph{team performance} and individual \emph{job satisfaction} are predicted by both \emph{psychological safety} and \emph{team norm clarity}. In more detail, our findings reveal that, for our data set, \emph{team norm clarity} is a stronger predictor of both concepts than \emph{psychological safety} (see the Beta values in Table~\ref{table:Regression_teameffectivness} and Table~\ref{table:Regression_jobsatisfaction}, respectively).

Our findings thus emphasize the importance of having clear norms in software development teams. That clarity of team norms was a stronger predictor than psychological safety and that it did not moderate the effects of psychological safety, indicate that clarity of team norms is a significant concept in its own and that both concepts are important for further study as well as for interventions. Previous research on norms has, mostly, focused on the content of norms determining the performance of teams~\cite{ellemers2005identity}. Our result instead highlights the importance of adopting norms that the team members are aware of and understand. Holding distinct norms imply that team members recognize what behavior that is accepted by the team, but it also suggests that team members know what behavior they can expect from their teammates. Such awareness arguably creates a more predictable psycho-social work environment that reduces uncertainties.

Moreover, by empirically linking psychological safety with both team performance and job satisfaction, our study reinforces the significance of psychological safety for the software engineering domain. Thereby, it also conforms to the large body of research postulating the concept's relevance in teamwork~\cite{sanner2013psychological, edmondson2014psychological, newman2017psychological,frazier2017psychological}. That is important since it opens up our field to the many studies and specific methods and interventions with which managers and team members can try to improve psychological safety in their teams and organization. 



Compared to the most similar study, of Faraj and Yan~\cite{faraj2009boundary}, we focus on agile teams in present-day software development organizations and study psychological safety and norm clarity directly rather as mediating factors in a particular type of team activity (`boundary work'). It is not clear that the results of Faraj and Yan on psychological safety, they did not study norm clarity, apply to the agile teams we study or what boundary work means in their context. Our work is also more general in that it points to our two independent variables being important for team performance regardless of the type of work being investigated. This also puts us in contrast to the work of Safdar et al~\cite{safdar2017can} which study knowledge sourcing and not team activities as a whole. Besides, they also only study psychological safety.

Taken together, results from our study provide initial directions for software engineering companies to follow when aiming to improve team and organizational performance. We believe that the simplicity of our findings, from a practitioner's point-of-view, has advantages. Previous research has identified a vast amount of concepts that affect team performance~\cite{kozlowski2006enhancing}. Our results help software engineering organizations to focus their efforts on identifying essential concepts to consider primarily. Still, we reiterate that the factors affecting any team's performance are likely to be complex and larger studies including more factors and relating their effects in more detail are needed.

Finally, our efforts have provided promising yet initial results in an important research area. Given the relevance of teamwork in software organizations and the significant effects of norms on team performance, we think that it is high time that social norms are further, and more profoundly, explored in the software engineering context.

\subsection{Limitations}
First, we acknowledge that our first-order~\footnote{The order of approximation indicates how precise an approximation is. First-order approximation is the term used for a further educated guess at an answer~\cite{wiki:ordersofapproximation}.} model, which included four variables, cannot be considered complete. Instead, it is an initial approximation that captures some of the most significant effects based on existing research in software engineering and on team performance in the social sciences. Studying more factors and their interaction will be needed to get a more in-depth understanding.

Even if we strove to use concepts that have previously been proven in research, we recognize a threat to validity. In particular, the representativeness of the measurement \emph{team performance}, which we estimated with self-assessment, can be questioned. A possible solution that would raise the validity would be to triangulate the data, i.e. cross-check the data using input collected from other sources. For example, objective measurement of team performance or rating by relevant managers, could help strengthen trust in validity of the findings. Still, since we reuse items and scales that have been previously used in the social sciences, we have analyzed the underlying items, and estimated the internal consistency, we believe that we can justify the use of the concept in the study and thus rate this threat as moderate/acceptable.

Finally, in this study, we used multiple linear regression to analyze the collected data. We recognize that using second generation data analysis techniques, such as partial least squares path analysis or LISREL, and/or Bayesian analysis, would have been viable options. Nonetheless, since our data passed the quality conditions for performing regression analysis, we do not consider our choice as a significant threat.

\section{Conclusion}
\label{sec_conclusion}
We have, in this study, demonstrated that psychological safety and clarity of team norms both affect software development teams' performance and job satisfaction. The analysis suggests that team norm clarity has a more significant impact than psychological safety.

Our findings thus emphasize the importance of adopting and clarifying distinct norms which contribute to a psycho-social work environment that reduces uncertainty. They also highlight the need to examine the relationship between social norms and software development more thoroughly.

\section{Acknowledgments}
We acknowledge the support of Swedish Armed Forces, Swedish Defense Materiel Administration and Swedish Governmental Agency for Innovation Systems (VINNOVA) in the project `Team-based development for safety and quality of avionic software' (project number 2017-04874).
%
\bibliographystyle{abbrv}
\bibliography{lenberg_ref}

%
%
\end{document}